\pdfoutput=1
\documentclass[aps,prd,reprint,twocolumn,showpacs,superscriptaddress,nofootinbib]{revtex4-1}  
\usepackage{tabularx}
\makeatletter
\def\hlinewd#1{%
\noalign{\ifnum0=`}\fi\hrule \@height #1 %
\futurelet\reserved@a\@xhline}
\makeatother
\usepackage{scrextend}
\usepackage{amsmath}
\usepackage{amssymb}
\usepackage{epsfig}
\usepackage{graphicx}
\usepackage{hyperref}
\usepackage{dcolumn}   
\usepackage{slashed}
\usepackage{color}
\usepackage{rotating}
\usepackage[margin=0.9in,a4paper]{geometry}
\usepackage[table,xcdraw,dvipsnames]{xcolor}
\usepackage[utf8]{inputenc}
\usepackage{colortbl}
\usepackage[normalem]{ulem}
\usepackage{mathrsfs} 
\usepackage[version=4]{mhchem}
\usepackage{cancel}

\definecolor{nicered}{rgb}{0.7,0.1,0.1}
\definecolor{nicegreen}{rgb}{0.1,0.5,0.1}
\definecolor{red}{rgb}{1.0, 0, 0}

\newcommand{\A}{{\cal A}}

\newcommand{\X}{{\cal X}}

\newcommand{\SU}{{\rm SU}}

\newcommand{\U}{{\rm U}}

\newcommand{\M}{{\cal M}}
\renewcommand{\L}{{\cal L}}
\newcommand{\N}{{\cal N}}
\newcommand{\E}{{\cal E}}
\newcommand{\Y}{{\cal Y}}

\renewcommand{\(}{\left(}
\renewcommand{\)}{\right)}

\newcommand{\bdm}{\begin{displaymath}}
\newcommand{\edm}{\end{displaymath}}
\newcommand{\bea}{\begin{eqnarray}}
\newcommand{\eea}{\end{eqnarray}}

\newcommand\T{\mathcal T}

\renewcommand{\S}{\mathcal{S}}
\renewcommand{\X}{\mathcal{X}}

\def\be{\begin{equation}}
\def\ee{\end{equation}}

\definecolor{nicered}{rgb}{0.7,0.1,0.1}
\definecolor{nicegreen}{rgb}{0.1,0.5,0.1}
\definecolor{red}{rgb}{1.0, 0, 0}
\definecolor{niceblue}{rgb}{0,0,0.8}
\definecolor{red}{rgb}{1.0, 0, 0}
\hypersetup{colorlinks,citecolor= nicegreen,linkcolor= nicered,urlcolor=nicered}

\def\eq#1{{Eq.~(\ref{#1})}}
\def\eqs#1#2{{Eqs.~(\ref{#1})--(\ref{#2})}}

\def\fig#1{{Fig.~\ref{#1}}}

\def\Table#1{{Table~\ref{#1}}}

\def\sect#1{{Sect.~\ref{#1}}}

\def\vev#1{\left\langle #1\right\rangle}

\def\gsim{\raise0.3ex\hbox{$\;>$\kern-0.75em\raise-1.1ex\hbox{$\sim\;$}}}
\def\lsim{\raise0.3ex\hbox{$\;<$\kern-0.75em\raise-1.1ex\hbox{$\sim\;$}}}

\def\mb[#1]{\mathbf{#1}}

\renewcommand{\bar}{\overline}

\definecolor{LightCyan}{rgb}{0.88,1,1}
\definecolor{piggypink}{rgb}{0.99, 0.87, 0.9}
\definecolor{applegreen}{rgb}{0.55, 0.71, 0.0}
\definecolor{darkpastelgreen}{rgb}{0.01, 0.75, 0.24}
\definecolor{green-yellow}{rgb}{0.68, 1.0, 0.18}

\newcommand{\beq}{\begin{equation}}
\newcommand{\eeq}{\end{equation}}
\newcommand{\beqa}{\begin{eqnarray}}
\newcommand{\eeqa}{\end{eqnarray}}

\newcommand{\eqn}[1]{Eq.~(\ref{#1})}

\begin{document}



\title{Gauged $\tau$-lepton chiral currents and $B \to K^{(*)} E_{\rm miss}$}

\author{Luca Di Luzio}
\email{luca.diluzio@pd.infn.it}
\affiliation{\small \it Istituto Nazionale di Fisica Nucleare, Sezione di Padova, Via F.~Marzolo 8, 35131 Padova, Italy}

\author{Marco Nardecchia}
\email{marco.nardecchia@roma1.infn.it}
\affiliation{\small \it Università degli Studi di Roma La Sapienza and INFN Section of Roma 1, Piazzale Aldo Moro 5, 00185 Roma, Italy}

\author{Claudio Toni}
\email{claudio.toni@lapth.cnrs.fr}
\affiliation{\small \it LAPTh, Université Savoie Mont-Blanc et CNRS, 74941 Annecy, France}

\begin{abstract}
\noindent
We consider a class of theories with a $\U(1)_X$ gauge symmetry associated with leptonic chiral currents. The low-energy effective field theory includes a light spin-1 boson coupled to the electroweak gauge sector via a Wess–Zumino term, which ensures anomaly cancellation in the infrared. As a concrete application, we show that a light vector boson with mass $m_X \simeq 2.1\,\text{GeV}$, coupled to a $\tau$-lepton chiral current, can readily account for the recent $3\sigma$ excess observed in $B \to K^{(*)} E_{\rm miss}$ at Belle II, while remaining consistent with existing constraints from $Z \to \gamma E_{\rm miss}$ and direct searches for anomalon fields responsible for anomaly cancellation in the ultraviolet. 
After classifying phenomenologically viable models, we explore 
in greater detail two concrete realizations which give rise to distinctive phenomenological signatures, potentially accessible at future experiments at the high-energy and intensity frontiers.

\end{abstract}

\maketitle


\section{Introduction}
\label{sec:intro}

The physics of light dark sectors has witnessed a growing amount of interest in recent years, driven by both theoretical developments and experimental advances. These sectors, often populated by particles with sub-GeV masses and feeble interactions with the Standard Model (SM), offer 
attractive alternatives to traditional dark matter scenarios and can provide explanations for longstanding anomalies in particle physics and cosmology. A wide variety of experimental efforts, from high-intensity beam-dump experiments  
to low-energy precision 
measurements and astrophysical observations, are actively probing this elusive territory. On the theoretical side, models involving dark photons, axion-like particles, and other light mediators have led to a rich phenomenology, especially when embedded in frameworks that address broader questions such as the strong CP problem, the hierarchy problem or the 
nature of dark matter. 

In many of these scenarios, the ultraviolet (UV) completion of the model plays a crucial role: not only does it ensure theoretical consistency, but it can also make the framework more predictive. A notable example is provided by light spin-1 bosons coupled, at low energies, to fermionic currents that are anomalous under the electroweak (EW) group.  
Differently from the case of a secluded abelian gauge boson kinetically mixed with the photon~\cite{Holdom:1985ag}, and hence universally coupled to the SM sector via the electromagnetic current, or the gauging of an anomaly-free linear combination of SM accidental symmetries, such as $B/3 - L_i$, these scenarios generally require the introduction of new fermions, often referred to as \emph{anomalons}, which cancel the anomalies of the new $\U(1)_X$ gauge symmetry, including mixed anomalies with the EW group. 

These anomalons must carry EW charges and, in order to evade detection at high-energy colliders, need to be heavier than the EW scale. As a result, their effects on the low-energy physics of the light vector boson associated with the $\U(1)_X$ gauge symmetry, denoted here as $X$, can be captured within an effective field theory (EFT) framework. In particular, integrating out the heavy fermions at one loop generates dimension-4 Wess-Zumino (WZ) terms \cite{Wess:1971yu,Witten:1983tw}, schematically of the form $X (W \partial W + W W W)$ and $X B \partial B$, where $W$ and $B$ denote the $\SU(2)_L$ and $\U(1)_Y$ gauge bosons, respectively. 
These contact interactions play the role of compensating, within the EFT that no longer contains the anomalon fields, the anomalous variation of the effective action induced by the SM fermion current coupled to $X$ (see \emph{e.g.}~\cite{DHoker:1984izu,DHoker:1984mif,Preskill:1990fr,Feruglio:1992fp}).

As emphasized in Refs.~\cite{Dror:2017ehi,Dror:2017nsg}, WZ terms exhibit an axion-like behavior, which can be understood via the equivalence theorem applied to the longitudinal component of $X$, and they lead to amplitudes that grow with energy. The anomalous $X W \partial W$ vertex can be dressed with SM flavor-violating interactions, giving rise to loop-induced flavor-changing neutral current 
processes such as $B \to K X$ and $K \to \pi X$,  
following a predictive flavor pattern 
which realizes the so-called minimal flavor violation hypothesis (MFV) \cite{DAmbrosio:2002vsn}. 
Meanwhile, the anomalous $X B \partial B$ vertex induces the decay $Z \to \gamma X$ at tree level (see also~\cite{Ismail:2017fgq,Michaels:2020fzj,Davighi:2021oel,Kribs:2022gri}). In both cases, these processes are enhanced as $(\text{energy} / m_X)^2$, and therefore they typically lead to the most stringent constraints on light vectors that do not couple directly to electrons.

Ref.~\cite{DiLuzio:2022ziu} explored the possibility of suppressing the coefficients of the aforementioned WZ terms by means of a UV completion in which the mass of the anomalons arises entirely from EW-symmetry-breaking sources. In this setup, the anomalous couplings of the longitudinal component of $X$ to the EW gauge bosons vanish, thereby relaxing the otherwise stringent bounds on light vectors. This scenario is particularly constrained by measurements of the $h \to Z \gamma$ decay rate and direct searches for non-decoupling charged leptons. A more recent analysis \cite{Barducci:2023zml} demonstrated that such scenarios are, in fact, perturbatively excluded by current LHC limits on new chiral charged leptons.

Conversely, in the phenomenologically viable 
case where the mass of the anomalons 
stems from the vacuum expectation value (VEV) of a SM-singlet field, 
the low-energy coefficients of the WZ terms are entirely 
fixed by the requirement of cancelling the $\SU(2)_L^2\U(1)_X$ and $\U(1)_Y^2\U(1)_X$ anomalies of the SM  
sector (see \emph{e.g.}~\cite{Dror:2017nsg}), 
thus providing a particularly predictive setup.  

While previous analyses~\cite{Dror:2017ehi,Dror:2017nsg,DiLuzio:2022ziu} focused on the possibility of gauging a linear combination of the SM accidental symmetries (\emph{i.e.}~baryon and family-lepton numbers), a choice that imposes non-trivial constraints on the structure of the WZ terms, in this work, we extend those setups by considering the gauging of a chiral component of family-lepton number, characterized by gauge charges that differ, in general, between left- and right-handed leptons. This framework allows for greater flexibility in the structure of the WZ terms, but it also enforces certain texture zeros in the charged-lepton Yukawa matrix. These zeros must be lifted via higher-dimensional operators in the spirit of the Froggatt-Nielsen (FN) 
mechanism~\cite{Froggatt:1978nt,Leurer:1992wg,Leurer:1993gy} in order to generate realistic charged lepton masses.

As an application of this setup, we present a light new physics interpretation of the recent evidence for the decay mode $B^+ \to K^+ E_{\rm miss}$ reported by the Belle II collaboration~\cite{Belle-II:2023esi}, which shows a $2.9\sigma$ excess over the SM prediction~\cite{Parrott:2022zte,Becirevic:2023aov}, 
triggering extensive phenomenological investigations \cite{Athron:2023hmz,Allwicher:2023xba,Felkl:2023ayn,Wang:2023trd,He:2023bnk,Chen:2023wpb,Berezhnoy:2023rxx,Datta:2023iln,Altmannshofer:2023hkn,McKeen:2023uzo,Fridell:2023ssf,Ho:2024cwk,Loparco:2024olo,Gabrielli:2024wys,Hou:2024vyw,Chen:2024cll,He:2024iju,Bolton:2024egx,DAlise:2024qmp,Marzocca:2024hua,Becirevic:2024pni,Buras:2024ewl,Kim:2024tsm,Rosauro-Alcaraz:2024mvx,Hati:2024ppg,Allwicher:2024ncl,Becirevic:2024iyi,Altmannshofer:2024kxb,Guedes:2024vuf,Buras:2024mnq,Bhattacharya:2024clv,Hu:2024mgf,Lee:2025jky,Lin:2025jzp,Calibbi:2025rpx,Berezhnoy:2025tiw,Bolton:2025fsq,Aliev:2025hyp,Ding:2025eqq}. 
In particular, it has been shown that the excess can be explained by a two-body decay involving an invisible particle with mass
$m_X=(2.1\pm0.1)$ GeV \cite{Altmannshofer:2023hkn,Bolton:2024egx,Bolton:2025fsq}. 
In our setup, this is naturally obtained from gauging a chiral component of the $\tau$-lepton current, with a non-vanishing projection onto the left-handed component. As a result, the dominant decay channel of the $X$ boson is into neutrinos (\emph{i.e.}~missing energy). The coupling to $b$-$s$ quarks originates from the anomalous $X W \partial W$ vertex, dressed at one loop by SM flavor-violating interactions.

The paper is structured as follows. In \sect{sec:EFTWZ}, we develop the EFT of a light gauge boson coupled to an anomalous current, extending the analysis to non-vectorial currents. We then interpret the processes $B^+ \to K^+ X$ and $Z \to \gamma X$ within this EFT framework. 
In \sect{sec:axial}, we construct a minimal anomaly-free setup based on gauging the chiral $\tau$-lepton number, which requires the introduction of a suitable set of anomalon fields. Next, we identify phenomenologically viable models capable of explaining the $B^+ \to K^+ E_{\rm miss}$ excess.
Two concrete realizations are explored in detail in \sect{sec:UVmodel}, with particular attention to reproducing the observed charged-lepton Yukawa textures. 
We summarize our results and conclusions in
\sect{sec:concl}.

\section{EFT of a light gauge boson 
coupled to an anomalous current} 
\label{sec:EFTWZ}

Consider a spontaneously broken $\U(1)_X$ gauge symmetry, 
featuring a light gauge boson, $X$, with $m_X \ll v \simeq 246$ GeV.  
At low energies, it couples to a SM fermionic current
\beq 
\label{eq:LU1X}
\mathscr{L}_X \supset -g_X X_\mu J_{X,\text{SM}}^{\mu} \, ,
\eeq
with
\be
\label{eq:defJXSM}
J_{X,\text{SM}}^{\mu}=\sum_{f\in\text{SM}} \alpha_f \bar{f}\gamma^\mu f \ ,
\ee
and the sum is performed over the chiral SM fields $f=\{q_L^i,u_R^i,d_R^i,\ell_L^i,e_R^i\}$.
Our convention for the covariant derivative is $D_\mu \equiv \partial_\mu+ig_X \X X_\mu$, with $\X$ denoting the $\U(1)_X$ charge, and similarly for the SM gauge bosons. 
We emphasize that the fermionic fields appearing at low energies, labeled above as SM fields, are not necessarily $\U(1)_X$ eigenstates. That is, in general, $\alpha_f \neq \X_f$, since the low-energy states may arise as linear combinations of UV fields with different $\U(1)_X$ charges (this is \emph{e.g.} the case of the model considered in \sect{sec:Lm4Rmodel}).

The presence of a new vector field coupled to the fermions induces an anomalous breaking of the gauge currents, 
despite the fact that the Lagrangian term ${\mathscr L}_X$ being invariant under the SM symmetries.
In a path integral approach, this is understood as due to the non-invariance of the functional measure \cite{Fujikawa:1980eg}.

As an example, assuming a regularization scheme that treats symmetrically all the gauge fields,\footnote{This choice is denoted as ``consistent anomaly'', opposite to the ``covariant anomaly'' one where the regularization scheme preserves some of the gauge symmetries. The coefficients of the anomaly and of the WZ terms are then regularization-scheme dependent. This ambiguity however vanishes when the contribution from triangular diagrams and WZ terms is summed together.  
} the anomalous breaking of the hypercharge current is found to be
\be
\label{eq:divcurY}
\partial_\mu J_{Y,\text{SM}}^\mu=g_{X}g'\frac{{\cal A}_{XYY}^\text{SM}}{24\pi^{2}} \epsilon^{\alpha\mu\beta\nu}(\partial_{\alpha}X_{\mu})(\partial_{\beta}B_{\nu}) \ ,
\ee
where $\epsilon^{0123}=+1$ and 
we have introduced the anomaly traces defined as
\be
\label{eq:defAnom}
\! {\cal A}_{\alpha\beta\gamma}\equiv\text{Tr}\left[T_\alpha \{T_\beta,T_\gamma\}\right]\Big|_\text{R} - \text{Tr}\left[T_\alpha \{T_\beta,T_\gamma\}\right]\Big|_\text{L} \, ,
\ee
with $T_{\alpha}$ being any generator of the gauge group.
Similarly, one finds
\be
\label{eq:divcurX}
\begin{split}
\partial_\mu J_{X,\text{SM}}^\mu=&g'^2\frac{{\cal A}_{XYY}^\text{SM}}{48\pi^{2}} \epsilon^{\alpha\mu\beta\nu}(\partial_{\alpha}B_{\mu})(\partial_{\beta}B_{\nu}) \\
+& \text{ explicit breaking terms} \, ,
\end{split}
\ee
where we have included possible explicit breaking terms arising from fermion mass terms in the $\U(1)_X$-broken phase, which are present if the $\U(1)_X$ symmetry has an axial component not aligned to $\U(1)_Y$.

The EW anomalies are 
cured in the UV by the presence of anomalon fields,  
which are chiral under the $\U(1)_X$ 
and transform non-trivially under the EW group.
In the low-energy EFT instead, below the $\U(1)_X$-breaking scale, the EW anomalies are compensated by effective operators which appear upon integrating out the anomalon fields at one loop. 
Assuming the phenomenologically motivated case where 
the mass of the anomalons stems from the VEV of a SM-singlet field,
one finds \cite{DHoker:1984izu,DHoker:1984mif,Dror:2017ehi,Dror:2017nsg,DiLuzio:2022ziu} 
\begin{align}
\label{EFT}
\Gamma_{\rm WZ} &= \int d^4 x \ {\cal L}_{\rm WZ} \nonumber\\
&=\int d^4 x \Bigg[
g_{X}g'^{2}\frac{C_{BB}}{24\pi^{2}}\epsilon^{\alpha\mu\nu\beta}X_{\alpha}B_{\mu}\partial_{\beta}B_{\nu} \nonumber \\
&+g_{X}g^{2}\frac{C_{WW}}{24\pi^{2}}\epsilon^{\alpha\mu\nu\beta}X_{\alpha}W_{\mu}^{a}\partial_{\beta}W_{\nu}^{a} \nonumber \\
&+\text{non-abelian terms}\Bigg]\,,  
\end{align}
where $a = 1,2,3$ and we neglected non-abelian $W$ terms scaling with an extra gauge coupling $g$.

The requirement of having an anomaly-free theory
relates the WZ coefficients to the anomalies induced solely by the SM fields as
\be
\label{eq:trace1}
C_{WW}=-{\cal A}_{XWW}^\text{SM} \,, \quad C_{BB}= -{\cal A}_{XYY}^\text{SM} 
\ .
\ee
As a consistency check, under a local gauge variation of the $\U(1)_Y$ symmetry, the hypercharge field transforms as
\be
\delta_\alpha B_\mu=\partial_\mu\alpha(x) \, ,
\ee
yielding after integrating by parts
\begin{align}
&\delta_\alpha\Gamma_{\rm WZ}=\int d^4 x \nonumber \\
&\times \Bigg[
g_{X}g'^{2}\frac{{\cal A}_{XBB}^\text{SM}}{24\pi^{2}}\alpha(x) \, \epsilon^{\alpha\mu\beta\nu}(\partial_{\alpha}X_{\mu})(\partial_{\beta}B_{\nu})\nonumber \Bigg]\,,
\end{align}
which cancels the anomalous breaking of the current in \eq{eq:divcurY}.

We note that in Ref.~\cite{DiLuzio:2022ziu}, the condition
\beq 
{\cal A}_{XAA}^\text{SM} \equiv {\cal A}_{XWW}^\text{SM} + {\cal A}_{XYY}^\text{SM} = 0 
\eeq
was imposed, which corresponds in the UV theory to the gauging of a purely vectorial SM current. In this work, we extend the framework to also allow for the gauging of an axial component of the SM current. The price to pay is a tree-level breaking of the current coupled to $X$, induced by the masses of the SM fermions. This leads to enhancements of order $(m_\text{SM}/m_X)^2$ in certain processes, which may be phenomenologically relevant (see \emph{e.g.}~\cite{Kahn:2016vjr,DiLuzio:2023xyi}). 

Finally, more general values of the WZ coefficients, 
$C_{WW}$ and $C_{BB}$, are permitted if the anomalon mass acquires a component that breaks the SM symmetry, with general formulae provided in Appendix~A of Ref.~\cite{DiLuzio:2022ziu}.

\subsection{$B \to K^{(*)} X$ and $Z \to \gamma X$}
\label{sec:BKXEFT}

\begin{figure*}[t!]
  \centering
  \includegraphics[width=1.00\textwidth]{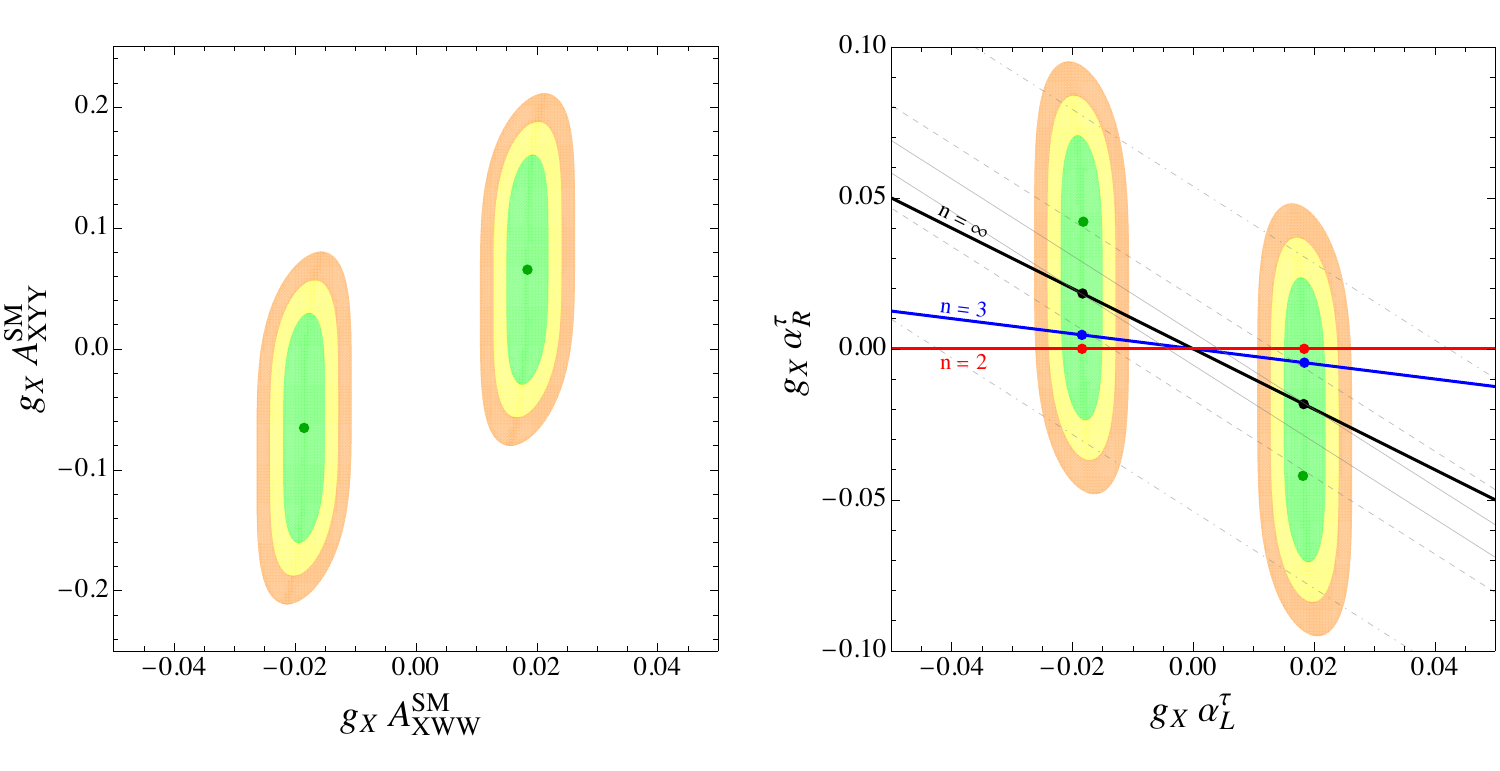}
  \caption{The green, yellow, and orange bands denote the 1$\sigma$, 2$\sigma$, and 3$\sigma$ regions 
favored by the $B \to K^{(*)} X$ and $Z \to \gamma X$ observables, shown in the plane of anomaly traces (left panel) and $\tau$-lepton $\mathrm{U}(1)_X$ charges (right panel). 
Two examples of correlations are displayed in the 
right panel, corresponding to the $L$-only model ($n=2$, in red) and the $L=-4R$ model ($n=3$, in blue). The blue and red points refer to the best-fit of each model.
We also show the asymptotic case $n=\infty$ in black. The gray lines correspond to ${\mathcal B}(Z\to\gamma X)=10^{-8}\text{ (solid)},10^{-7}\text{ (dashed)},10^{-6}\text{ (dot-dashed)}$.
}
  \label{fig:plot}
\end{figure*}

In the gaugeless limit, \emph{i.e.}~$g_X\to0$ while $m_X/g_X$ fixed,
the longitudinal modes of the $X$ field dominate the physical amplitudes. Through the usual Goldstone boson equivalence theorem, we can isolate the longitudinal mode in this limit via the substitution $X_\mu\to\partial_\mu\xi/m_X$, 
where $\xi$ denotes the Goldstone field. Applying this to \eq{eq:LU1X} and (\ref{EFT}) and integrating by parts, while keeping in mind \eq{eq:divcurX}, it leads to the axion-like operator
\begin{align}
\label{eq:LXplusLWZ}
&{\cal L}_X+{\cal L}_{\rm WZ} \supset \nonumber \\ 
&g_{X}g^{2}\frac{{\cal A}_{XWW}^\text{SM}}{16\pi^{2}}\frac{\xi}{m_X} \, \epsilon^{\alpha\mu\beta\nu}(\partial_{\alpha}W^a_{\mu})(\partial_{\beta}W^a_{\nu}) \, .
\end{align}
Upon integrating out the $W$ boson at the one-loop level, 
the axion-like operator $\xi W^{-}\tilde{W}^{+}$ yields the effective interaction 
\beq
\label{EFT FCNC}
g_{\xi d_{i}d_{j}}\bar{d}_{j}\gamma^{\mu}P_{L}d_{i}\,(\partial_{\mu}\xi/m_{X})\,+\text{h.c.} \, , 
\eeq
in terms of the effective coupling \cite{Izaguirre:2016dfi}
\beq 
\label{eq:gxididj}
g_{\xi d_{i}d_{j}}=-\frac{3g_{X}g^{4}}{(4\pi)^{4}}{\cal A}_{XWW}^\text{SM}\sum_{\alpha=u,c,t}V_{\alpha i}V_{\alpha j}^{*}F(m_{\alpha}^{2}/m_{W}^{2}) \, ,
\eeq
with $V$ denoting the CKM matrix and the loop function
\beq 
F(x)=\frac{x(1+x(\ln x-1))}{(1-x)^{2}} \, .
\eeq
This leads to $B\to K^{(*)} X$ decays which are investigated at flavor factories. Interestingly, the Belle II collaboration has recently observed a $2.9\sigma$ excess over the SM in the decay mode $B^+ \to K^+ E_\text{miss}$, 
while no evidence for $B \to K^{*} E_\text{miss}$ was found 
at BaBar \cite{BaBar:2013npw}.
In particular, it has been shown that this 
excess could be fit by a 
two-body decay involving an invisible state with mass $m_X=(2.1\pm0.1)$ GeV \cite{Altmannshofer:2023hkn,Bolton:2024egx,Bolton:2025fsq}.
Matching the couplings stemming from the WZ operators in \eq{eq:gxididj}\footnote{Note that the above expressions are obtained in the limit of validity of the equivalence theorem. At the best-fit point, corrections of order $m_X^2 / m_B^2 \approx 15\%$ are expected.} 
with the parametrization of Ref.~\cite{Bolton:2024egx,Bolton:2025fsq}, \emph{i.e.}
\be
\label{eq:defgSgP}
g_S\equiv-\frac{m_b-m_s}{2}\frac{g_{\xi sb}}{m_X} \ , \ \ \ g_P\equiv\frac{m_b+m_s}{2}\frac{g_{\xi sb}}{m_X} \,,
\ee
the analysis of \cite{Bolton:2024egx,Bolton:2025fsq} on the full $b\to s E_\text{miss}$ data found the best-fit values
\begin{align}
\label{eq:bfgS}
\left|g_S\right|&=(1.6\pm0.2)\times10^{-8} \, , \\
\label{eq:bfgP}
\left|g_P\right|&<2.5\times10^{-8} \text{ (at 90\% C.L.)} \, .
\end{align}
The current in \eq{EFT FCNC} induces other flavor violating decays such as $B\to\pi X,\rho X$.\footnote{Instead, the $X$ boson cannot be emitted on-shell in $K\to\pi E_\text{miss}$ decays and we checked that the off-shell effects lead to a negligible contribution for the expected value of the gauge coupling.}
The present bounds on these decays read at 90\% C.L.~\cite{Belle:2017oht}
\begin{align}
    \mathcal{B}(B^{+}\to\pi^{+}E_\text{miss}) &< 1.4\times10^{-5} \, , \\
    \mathcal{B}(B^{0}\to\pi^{0}E_\text{miss}) &< 0.9\times10^{-5} \, , \\ 
    \mathcal{B}(B^{+}\to \rho^{+}E_\text{miss}) &< 3.0\times10^{-5} \, , \\
    \mathcal{B}(B^{0}\to\rho^{0}E_\text{miss}) &< 4.0\times10^{-5} \, ,
\end{align}
which are well above the SM predictions \cite{Bause:2021cna}. 
Given the MFV structure of the interaction, 
at the best fit point one finds: 
\begin{align}
{\mathcal B}(B^+\to\pi^+ X)_\text{best-fit}&=2.7\times10^{-7} \, , \\
{\mathcal B}(B^0\to\pi^0 X)_\text{best-fit}&=1.2\times10^{-7} \, , \\
{\mathcal B}(B^+\to\rho^+ X)_\text{best-fit}&=2.7\times10^{-7} \, , \\
{\mathcal B}(B^0\to\rho^0 X)_\text{best-fit}&=1.3\times10^{-7} \, , 
\end{align} 
of the same order of the SM predictions \cite{Bause:2021cna}.
Similarly to \eq{eq:LXplusLWZ}, for the neutral sector 
one obtains the 
Lagrangian operator
\begin{align}
\label{eq:xidZdAlag}
&{\cal L}_X+{\cal L}_{\rm WZ} \supset c_{XZ\gamma} \frac{\xi}{m_{X}}\epsilon^{\alpha\mu\beta\nu}(\partial_{\alpha}Z_{\mu}^{a})(\partial_{\beta}A_{\nu}) \, , \nonumber \\ 
&c_{XZ\gamma}=g_{X}gg'\frac{c_W^2{\cal A}_{XWW}^\text{SM}-s_W^2{\cal A}_{XYY}^\text{SM}}{8\pi^{2}}  \, ,
\end{align}
with $s_W=\sqrt{1-c_W^2}=g'/\sqrt{g^2+g'^2}$,
which induces the decay $Z\to\gamma X$ with a rate given by \cite{Dror:2017ehi}
\begin{align}
\Gamma(Z\to\gamma X)&\approx\frac{|c_{XZ\gamma}|^2}{96 \pi}\frac{m_Z^3}{m_X^2} \, .
\end{align}
The strongest constraint on these decays comes from the L3 experiment at LEP and reads \cite{L3:1997exg}
\be
\label{eq:LEPbound}
{\mathcal B}(Z\to\gamma X) \lesssim 10^{-6} \text{  at 95\% C.L.} \ .
\ee
Expressing the constraints in terms of the anomaly traces in \eqn{eq:defAnom}, we construct a $\chi^2$ variable of $B\to K^{(*)}X$, $B\to\pi X,\rho X$ and $Z\to \gamma X$ observables. 
We note, however, that $B\to\pi X,\rho X$ decays 
give a negligible contribution to the fit.
We then plot the regions of the parameter space compatible with the experimental bounds at $1\sigma$, $2\sigma$ and $3\sigma$ in the left panel of Fig.~\ref{fig:plot}.
In particular, the allowed 1$\sigma$ range for 
the ratio of the anomaly traces is ${\cal A}_{XYY}^\text{SM}/{\cal A}_{XWW}^\text{SM} \in [-1.7,8.8]$.

\section{Gauging chiral $\tau$-lepton number} 
\label{sec:axial}

To obtain the above-mentioned EFT setup 
we consider a scenario with gauged $\tau$-lepton number, 
keeping in general $\alpha^\tau_L$ and $\alpha^\tau_R$ 
uncorrelated
in \eq{eq:defJXSM} and set to zero all the other 
coefficients. This choice is motivated by the need to tame extra constraints from current interactions, which are especially severe \emph{e.g.}~in the case of couplings to electrons. 
Another reason for coupling $X$ to $\tau$ is that, 
for the $m_X = 2.1$ GeV benchmark, $X \to \tau\tau$ is kinematically closed. So the main decay channel of $X$ is into invisible final states (tau neutrinos), as required by $B \to K^{(*)} E_{\rm miss}$.\footnote{Note that the $X \to \gamma \gamma$ decay is forbidden by the Landau-Yang theorem \cite{Landau:1948kw,Yang:1950rg}. That would not be the case with an axion-like particle coupled to $W \tilde W$ and $B \tilde B$, which would dominantly decay into photons, unless the coupling to photons is tuned away.} 

The strongest constraints on axial-vector couplings of the $X$ boson to $\tau^+ \tau^-$ arise from $\tau^+ \tau^- X$ production at $B$ factories and tau/charm factories~\cite{Dror:2017nsg}. If $X$ couples axially to $\tau$ leptons, the process $e^+ e^- \to \tau^+ \tau^- X$ receives an enhancement proportional to $(m_\tau / m_X)^2$ due to the emission of a longitudinally polarized $X$. 
Searches by BaBar and Belle are sensitive to this effect and constrain the combination $|\alpha_R^\tau - \alpha_L^\tau| g_X / (2m_X) \lesssim (150\,\text{GeV})^{-1}$ for $m_X \sim 2$~GeV \cite{Dror:2017nsg}, 
under the assumption that the $X$ boson decays leptonically into
muons and/or electrons (see, \emph{e.g.}, \cite{McKeen:2011aa,Batell:2016ove}). 
However, in our setup, the previous bound does not apply, as the $X$ boson decays into invisible final states, for which existing limits are expected to be weaker. 

Following the definition of the anomaly traces in \eq{eq:defAnom}, 
restricted to the light SM fields in the SM+$X$ EFT, 
we find\footnote{This expression assumes that the $\tau$-lepton $\U(1)_X$-current eigenstate is also a mass eigenstate. In general, as discussed in \sect{sec:Lm4Rmodel}, that might not be the case and extra contributions to the anomaly traces should be included in the SM+$X$ EFT.}
\begin{align}
\label{eq:AXYYmatch}
\A^{\rm SM}_{XYY} &=2 \alpha_R^\tau - \alpha_L^\tau \, , \\
\label{eq:AXWWmatch}
\A^{\rm SM}_{XWW} &=- \alpha_L^\tau \, .
\end{align}
Given the matching conditions above, 
the parameter space in the left panel of \fig{fig:plot} 
is mapped into the right panel of the same figure. 
Note that, in the chiral basis, 
$B \to K^{(*)} X$ and $Z\to\gamma X$, are respectively proportional to $\alpha_L^\tau$ and $\alpha_R^\tau$.

\subsection{Minimal anomalon setup}

The minimal setup of anomaly-canceling fermions 
for the gauging of $\tau$-lepton number 
is given in 
\Table{tab:genU1fieldcontent}, where, in addition to the anomalon fields, 
we also extend
the scalar sector of the SM in order to spontaneously break the $\U(1)_X$ symmetry. 
Note that we have included $N$ copies of chiral SM-singlet fermions $\N^{\alpha}_{R}$ 
($\alpha = 1, \ldots, N$)
which play a role for the cancellation of the 
$\U(1)_X$ and $\U(1)^3_X$ anomalies 
(the minimal choice being $N=2$), 
but whose presence does not impact the calculation of the 
WZ terms in the EW sector.

\begin{table}[h!]
	\centering
	\begin{tabular}{|c|c|c|c|c|c|}
	\hline
	Field & Lorentz & $\SU(3)_C$ & $\SU(2)_L$ & $\U(1)_Y$ & $\U(1)_{X}$ \\ 
	\hline 
	$\L_L$ & $(\tfrac{1}{2},0)$ & 1 & $n$ & $\Y_{\L}$ & $\X_{\L_L}$ \\
	$\L_R$ & $(0,\tfrac{1}{2})$ & 1 & $n$ & $\Y_{\L}$ & $\X_{\L_R}$ \\
	$\N_R^{\alpha}$ & $(0,\tfrac{1}{2})$ & 1 & 1 & $0$ & $\X^{\alpha}_{\N_R}$ \\
	\hline
    $\S$ & $(0,0)$ & 1 & 1 & 0 & $\X_{\S}$ \\  
	\hline
	\end{tabular}	
	\caption{\label{tab:genU1fieldcontent} 
	Anomaly-canceling field content for a $\SU(3)_C \times \SU(2)_L \times \U(1)_Y \times \U(1)_X$ gauge theory, 
	with $X$ acting as a chiral  
    $\tau$-lepton number. 
    An extended scalar sector is also included. 
    }
\end{table}

The $\U(1)_X$ charges are required to cancel all gauge anomalies. This corresponds to the following five conditions:
\begin{widetext}
\begin{align}
\label{eq:anom1}
\text{Gravity} \times \U(1)_X &:\quad n (\X_{\L_L} - \X_{\L_R}) - \sum_{\alpha} \X^{\alpha}_{\N_R} 
+ 2 \alpha^{\tau}_L - \alpha^{\tau}_R = 0 \, , \\
\label{eq:anom2}
\U(1)^3_X &:\quad n (\X_{\L_L}^3 - \X_{\L_R}^3) 
- \sum_{\alpha} \(\X^{\alpha}_{\N_R} \)^3 
+ 2  (\alpha^{\tau}_L)^3 -  (\alpha^{\tau}_R)^3
= 0 \, , \\
\label{eq:anom3}
\SU(2)^2_L \times \U(1)_X &:\quad  \frac{n(n^2 -1)}{12} (\X_{\L_L} - \X_{\L_R}) + \frac{1}{2}  \alpha^{\tau}_L = 0 \, , \\ 
\label{eq:anom4}
\U(1)^2_Y \times \U(1)_X &:\quad  n \Y_{\L}^2 (\X_{\L_L} - \X_{\L_R})   
+ \frac{1}{2}  \alpha^{\tau}_L -  \alpha^{\tau}_R = 0 \, , \\
\label{eq:anom5}
\U(1)_Y \times \U(1)^2_X &:\quad  n \Y_{\L} (\X_{\L_L}^2 - \X_{\L_R}^2)  
-  (\alpha^{\tau}_L)^2 +  (\alpha^{\tau}_R)^2
= 0 \, .
\end{align}
\end{widetext}
The first two conditions in \eqs{eq:anom1}{eq:anom2} 
fix two combinations of the $\U(1)_X$ 
charges of the $\N_R^\alpha$ field to be
\begin{align}
\label{eq:condU1Xn2}
\sum_{\alpha} \X^{\alpha}_{\N_R} &= n (\X_{\L_L} - \X_{\L_R}) 
+ 2  \alpha^{\tau}_L -  \alpha^{\tau}_R \, , \\
\label{eq:condU1Xn3}
\sum_{\alpha} \( \X^{\alpha}_{\N_R} \)^3 &= n (\X_{\L_L}^3 - \X_{\L_R}^3) + 2  (\alpha^{\tau}_L)^3 -  (\alpha^{\tau}_R)^3 \, . \nonumber    
\end{align}
Further constraints on the $\U(1)_X$ charges are obtained 
by the requirement that 
the anomalons 
pick up their mass from the VEV of $\S$, via the Yukawa interactions 
\beq 
\label{eq:yukVL}
- \Delta \mathscr{L}_{Y} = 
y_\L \bar{\L}_{L} \L_{R} \S^{*} + \text{h.c.} \, , 
\eeq 
which fixes 
\beq 
\label{eq:XScharges}
\X_\S = \X_{\L_R} - \X_{\L_L} \, .
\eeq
Renormalizable Yukawa interactions of the anomalons with the Higgs are forbidden by an appropriate choice of the gauge quantum numbers of $\L_{L,R}$, or, if necessary, by introducing additional discrete symmetries 
(see $L$-only model in \sect{sec:Lonlymodel}).

Putting together the remaining anomaly cancellation conditions, 
\eqs{eq:anom3}{eq:anom5},
and the invariance of the Yukawa sector, \eq{eq:XScharges}, 
we obtain the following solutions for the $\U(1)_X$ charges: 
\begin{align}
\label{eq:XsCharge}
\X_{\S} &= \frac{6}{n(n^2 - 1)} \alpha^{\tau}_{L}  \, , \\
\label{eq:YL}
\Y_{\L} &= \pm \sqrt{\frac{(n^2-1)(\alpha^{\tau}_L - 2\alpha^{\tau}_R)}{12 \alpha^{\tau}_L}} \, , \\ 
\label{eq:XLRp}
\X_{\L_L} + \X_{\L_R} &= \pm \sqrt{\frac{(n^2-1)}{3 \alpha^{\tau}_L (\alpha^{\tau}_L - 2\alpha^{\tau}_R)}} \nonumber \\  &\times ((\alpha^{\tau}_R)^2 - (\alpha^{\tau}_L)^2) \, . 
\end{align}
By adding a proper term in the scalar potential, $\Delta V (H, \S)$, 
the following VEV configurations 
are generated
\beq 
\vev{H} = \frac{1}{\sqrt{2}} 
\begin{pmatrix}
0 \\ v
\end{pmatrix} \, , \qquad 
\vev{\S} = \frac{v_X}{\sqrt{2}} \, , 
\eeq
with $v \approx 246$ GeV and $v_X$ being the order parameter of $\U(1)_X$ breaking. 
The latter is responsible for the mass of the $\U(1)_X$ gauge boson, $X^{\mu}$, that is
\beq 
\label{eq:massX}
m_{X} = \X_\S g_X v_X 
=  \frac{6}{n(n^2 - 1)} \alpha^{\tau}_{L} g_X v_X \, ,
\eeq
where $g_X$ is the $\U(1)_X$ gauge coupling entering the covariant derivative, 
\emph{i.e.}~$D^\mu \S \equiv (\partial^\mu + i g_X \X_\S X^{\mu}) \S$.

After $\U(1)_{X}$ symmetry breaking the Yukawa term in 
\eq{eq:yukVL}
gives mass 
to the anomalons
\begin{align} 
\label{eq:anomalonmass}
m_{\L} &=  
\frac{y_{\L}}{\sqrt{2}} v_X = 
\frac{y_{\L}}{\sqrt{2}} \frac{n(n^2 - 1)}{6} \frac{m_X}{\alpha^{\tau}_{L} g_X} \nonumber \\
&= 
\frac{y_{\L}}{\sqrt{2}} \frac{n(n^2 - 1)}{6} 120 \, \text{GeV} \( \frac{m_X}{2.1 \, \text{GeV}} \) \nonumber \\
&\times \( \frac{0.018}{{\alpha^{\tau}_{L} g_X}} \)
\, , 
\end{align}
where the expression above is normalized to the 
best-fit point $(m_X, \alpha^{\tau}_{L} g_X) = (2.1 \, \text{GeV}, 0.018)$, with the coupling determined from \eq{eq:bfgS}. 

The mass of the $\N_R^\alpha$ states is model-dependent, 
but they typically require the introduction of effective operators in the Yukawa sector (see \sect{sec:Lm4Rmodel}). 

We finally remark that, due to the chiral assignment of 
the $\U(1)_X$ $\tau$-charges, the $\tau$-Yukawa coupling, as well as the mixing with the other charged leptons, is absent at the 
renormalizable level. Hence, the model needs to be UV completed. This will be discussed in \sect{sec:UVmodel} 
for two specific realizations. 

\subsection{Scalar sector phenomenology}
\label{sec:Spheno}
While the main theoretical and phenomenological focus of this work is on the heavy anomalon fermions and the light gauge boson $X$, it is also important to outline the structure of the scalar sector.  

The spontaneous breaking of the $U(1)_X$ symmetry is achieved through the VEV of $\mathcal{S}$, whose value is fixed by fitting to the anomalous Belle~II data. In particular, from (\ref{eq:anomalonmass}) we obtain
\begin{equation}
v_X = \frac{n (n^2 -1)}{6}\, 120 \,\text{GeV}.
\end{equation}
In the explicit model considered in the next sections, we have $n=2$ and $n=3$, which lead to $v_X =120 \,\text{GeV}$ and $v_X =480 \,\text{GeV}$, respectively.  

This implies that the extended scalar sector contains two scalar fields: the SM Higgs $H$ and $\mathcal{S}$, which acquire VEVs of comparable size. In principle, they may exhibit significant mixing effects through the renormalizable operator $\lambda_{SH}\,\mathcal{S}^{\dagger} \mathcal{S}\, H^{\dagger} H$.  

The phenomenology of the scalar sector strongly depends on the value of $\lambda_{SH}$ and on the mass of the remaining radial mode of $\mathcal{S}$. These effects are generic and expected in all models with an extra gauged $U(1)$ symmetry, and we refer the reader to the extensive literature on this subject, see \emph{e.g.}~\cite{Dawson:2022zbb} and references therein.

Concerning Higgs decays, we note that loop-induced processes such as $h \to \gamma X$, $h \to ZX$, and $h \to XX$ are also present in our setup. We have explicitly checked that, in the parameter region favored by the Belle~II anomaly, these modes do not impose relevant constraints on the model. This conclusion is consistent with the findings of previous analyses in similar frameworks (see, 
\emph{e.g.}~Ref.~\cite{Bauer:2017ris}).

Finally, it is worth noting that a complete analysis of the bosonic sector should also include possible effects from kinetic mixing between the hypercharge and $X$ vector fields, a topic that has also been widely studied, see for example \cite{Fabbrichesi:2020wbt}.

\subsection{Anomalon searches}
\label{sec:anomalonpheno}

Given the best-fit point to explain $B \to K^{(*)} X$, 
the $\L$ anomalons tend to be rather light, as shown by \eq{eq:anomalonmass}. 
We discuss in turn the bounds on the anomalon masses, which depend on the quantum numbers of the $\L$ field and the charge of the lightest particle (LP) in the $n$-plet. 

In fact, if the LP in the $n$-plet is electrically charged and stable, it can provide striking signatures at colliders, in terms of charged tracks.
Typical mass bounds, \emph{e.g.}~from the ATLAS search~\cite{ATLAS:2023zxo},  
are at the level 
of $m_{\L} \gtrsim 1$ TeV, 
depending also on the electric charge of the LP 
(see \emph{e.g.}~\cite{DiLuzio:2015oha,Bonnefoy:2020gyh}). 
Similar bounds are obtained in the case when the LP in the $n$-plet mixes with SM leptons, 
by recasting same-sign lepton searches \cite{ATLAS:2017xqs}
(see \cite{Barducci:2023zml} for a recent analysis). 
In either case (stable charged LP or mixing with SM leptons), a somewhat large $n$ is required in order to evade those bounds 
(\emph{c.f.} \eq{eq:anomalonmass}). 

On the other hand, the lower bound on the anomalon masses 
is significantly reduced in the case that the LP is stable and neutral. Limits can either be set directly on the mass of the neutral particle as \emph{e.g.}~by mono-x
searches or from constraints on the invisible $Z$ width or, indirectly, by giving bounds on the
mass of the second lightest particle of the multiplet, such as in disappearing track signatures. Current limits at LHC, 
based on mono-x
searches and disappearing tracks, 
are still relatively weak, at the level of few 
hundreds of GeV, see \emph{e.g.}~\cite{Belyaev:2020wok}. 
The main reason being that the sensitivity of disappearing 
tracks is drastically reduced for a lifetime of the charged 
particle $\ll 10^{-10}$ sec. In fact, 
denoting with $-j \leq I \leq j$
the $T^3_L$ eigenvalue of the $n$-plet ($n=2j+1$), 
the decay rate of the $I$ component of the multiplet reads \cite{DiLuzio:2015oha} 
\begin{align}
\Gamma (\L_{I} \to \L_{I\pm1} \pi^\mp) &\approx \frac{(j\mp I)(j\pm I +1)}{7.5 \times 10^{-12} \, \text{sec}} \nonumber \\ 
&\times \( \frac{\Delta m}{500 \, \text{MeV}} \)^3 \, , 
\end{align}
that is valid for $m_{\L_I}>m_{\L_{I\pm1}}$ and $m_\L \gg\Delta m \gg m_{\pi^+}$, 
where $\Delta m$
is the radiative mass splitting of the 
charge eigenstates in the EW $n$-plet 
\cite{Cirelli:2005uq} 
\begin{align}
\label{eq:Deltam}
\Delta m  &\equiv m_{Q+1} - m_Q \approx 166 \, \text{MeV} \nonumber \\ 
&\times \(1 + 2Q + \frac{2 \Y_\L}{c_W} \) \, .
\end{align}
Hence, the cascade decays of the $n$-plet components are rather efficient, 
and disappearing tracks constraints are easily evaded in our setup. 
Also, LHC mono-jet searches
at 36 fb$^{-1}$ \cite{Belyaev:2020wok,ATLAS:2017bfj} are still relatively weak, 
\emph{e.g.}~at the level of 200 GeV for EW triplets. 

\subsection{Classifying phenomenologically viable models}

In the following, we further restrict the class of viable models by requiring 
that the LP in the EW $n$-plet 
is electrically neutral. This condition is motivated by the fact that current LHC constraints on EW $n$-plets 
with a neutral LP are particularly weak 
(\emph{cf.}~\sect{sec:anomalonpheno}), as well as by 
cosmological considerations, since the relic density of 
stable charged particles is highly constrained.

The requirement of a neutral component implies that $\Y_\L\in\{-j,-j+1,\dots,j-1,j\}$, hence $\Y_\L$ must be integer if $n$ odd or semi-integer if $n$ even.
Assuming $\Y_\L\geq0$, with no loss of generality,\footnote{If $\Y_\L<0$, one can redefine the EW anomalon field as $\L^\prime=\L^c$, where $c$ indicates charge conjugation.} 
one finds from \eq{eq:Deltam} that $m_{Q+1}-m_Q>0$ for $Q\geq0$, thus the positive charged components are heavier than the neutral one.
Regarding the negative charged components, let us assume that there is, at least, the $Q=-1$ component in the $n$-plet.
In this case, the requirement of a neutral LP reads $m_{Q=0}-m_{Q=-1}<0$, implying from \eq{eq:Deltam} $\Y_\L<c_W/2<1/2$ which admits only the integer or semi-integer positive solution $\Y_\L=0$.
However, we must discard this value because it does not lead to a viable solution of \eqs{eq:anom4}{eq:anom5}.
We then conclude that the lowest electric charge of the $n$-plet must vanish, which yields
\begin{align}
\Y_\L=&(n-1)/2 \, , \\
\alpha_R^\tau=&\frac{2-n}{1+n}\alpha_L^\tau \, , \\
\X_{\L_L} + \X_{\L_R}=&\frac{1-2n}{1+n}\alpha_L^\tau \, .
\end{align}
So, in particular, $-1\leq\alpha_R^\tau/\alpha_L^\tau\leq0$, and hence the possibility of a vectorial current,  
$\alpha_L^\tau = \alpha_R^\tau$, 
would require a different kind of UV completion \cite{WIP}. 
As an example, we show in \Table{tab:solfractional} the explicit value of the charges for $2\leq n\leq5$.

In summary, we are left with an infinite class of solutions 
labeled by $n$, with each of them able to explain the $B \to K^{(*)} E_{\rm miss}$ excess,  
while remaining consistent with $Z \to \gamma X$ 
at the $1\sigma$ level (see right panel of \fig{fig:plot}).

Note that large-$n$ representations suffer from perturbativity issues and, in addition, the 
$\U(1)_X$ charge assignments of the anomalons involve ratios of large integers, 
which are challenging to realize in explicit UV completions 
(cf.~\Table{tab:solfractional} for $n \geq 4$). 
We therefore focus on the two lowest-$n$ benchmark scenarios and construct explicit UV completions for these cases, 
also accounting for the charged-lepton mass structure.

\begin{table}[t!]
\centering
\begin{tabular}{|c|c|c|c|c|}
\hline
$n$ & $\Y_{\L}$ & $(\X_{\L_L} + \X_{\L_R})/\alpha^\tau_L$ & $\X_S/\alpha^\tau_L$ & $\X_{\L_R}/\X_{\L_L}$ \\
\hline
\rowcolor{piggypink}
$2$ & $1/2$ & $-1$ & $1$ & $0$ \\
\rowcolor{piggypink}
$3$ & $1$ & $-5/4$ & $1/4$ & $2/3$ \\
$4$ & $3/2$ & $-7/5$ & $1/10$ & $13/15$ \\
$5$ & $2$ & $-3/2$ & $1/20$ & $29/31$ \\
\vdots & \vdots & \vdots & \vdots & \vdots \\
\hline
\end{tabular}
\caption{Solutions of \eqs{eq:XsCharge}{eq:XLRp} 
featuring a neutral LP in the $n$-plet for $2\leq n\leq5$.
We highlight in pink the cases we consider in the next section.
}
\label{tab:solfractional}
\end{table}

\section{UV completions for $\tau$ mass generation}
\label{sec:UVmodel}

In this section, we present a UV completion for the two scenarios identified within the class of simplified models discussed in the previous section. A general feature of these constructions is the presence of vector-like fermions, which are required to generate a mass for the $\tau$ lepton via a FN mechanism~\cite{Froggatt:1978nt,Leurer:1992wg,Leurer:1993gy}.

\subsection{$L$-only model ($n=2$)}
\label{sec:Lonlymodel}

\begin{table}[!ht]
	\centering
	\begin{tabular}{|c|c|c|c|c|c|c|c|}
	\hline
	Field & Lorentz & $\SU(3)_C$ & $\SU(2)_L$ & $\U(1)_Y$ & $\U(1)_{X}$ & $Z_{2}^A$ & $Z_{2}^\tau$ \\ 
	\hline 
	$q^{i}_L$ & $(\tfrac{1}{2},0)$ & 3 & 2 & 1/6 & 0 & $+$ & $+$ \\ 
	$u^{i}_R$ & $(0,\tfrac{1}{2})$ & 3 & 1 & 2/3 & 0 & $+$ & $+$ \\ 
	$d^{i}_R$ & $(0,\tfrac{1}{2})$ & 3 & 1 & $-1/3$ & 0 & $+$ & $+$ \\ 
	$\ell^{1,2}_L$ & $(\tfrac{1}{2},0)$ & 1 & 2 & $-1/2$ & $0$ & $+$ & $+$ \\ 
	$e^{1,2}_R$ & $(0,\tfrac{1}{2})$ & 1 & 1 & $-1$ & $0$ & $+$ & $+$ \\ 
	$e^{3}_R$ & $(0,\tfrac{1}{2})$ & 1 & 1 & $-1$ & $0$ & $+$ & $-$ \\ 
    $\ell^{3}_L$ & $(\tfrac{1}{2},0)$ & 1 & 2 & $-1/2$ & $1$ & $+$ & $-$ \\ 
	$H$ & $(0,0)$ & 1 & 2 & 1/2 & $0$ & $+$ & $+$ \\
	\hline
	$\L_L$ & $(\tfrac{1}{2},0)$ & 1 & 2 & $-1/2$ & $0$ & $-$ & $+$ \\
	$\L_R$ & $(0,\tfrac{1}{2})$ & 1 & 2 & $-1/2$ & $1$ & $-$ & $+$ \\
	\hline
	$\E_L$ & $(\tfrac{1}{2},0)$ & 1 & 2 & $-1/2$ & $0$ & $+$ & $-$ \\
	$\E_R$ & $(0,\tfrac{1}{2})$ & 1 & 2 & $-1/2$ & $0$ & $+$ & $-$ \\
    \hline
	$\S$ & $(0,0)$ & 1 & 1 & 0 & $1$ & $+$ & $+$ \\  
	\hline
	\end{tabular}	
	\caption{\label{tab:Lm4Rmodel} 
	Field content of the $L$-only model.}
\end{table}

The field content of the model is summarized in \Table{tab:Lm4Rmodel}.
In this construction, gauge anomaly cancellation is achieved by introducing an $\SU(2)_L$ doublet anomalon ($n=2$), without the need for any SM-singlet fermions. This is possible because $\ell_L^3$ and $\L_R$, the only fermions charged under $\U(1)_X$, form a vector-like pair under the full gauge group.

To forbid an explicit mass term between $\ell_L^3$ and $\L_R$, we introduce a discrete $Z^A_2$ symmetry acting non-trivially only on the anomalon fields $\L_{L,R}$. This allows for a Yukawa interaction of the form given in \eq{eq:yukVL}. 
We further assume a separate discrete symmetry $Z_2^\tau$, interpreted as a $\tau$-flavor symmetry, which distinguishes $e_R^3$ from $e_R^{1,2}$. To generate the $\tau$ lepton mass via a FN mechanism, we introduce a vector-like fermion pair $\E_{L,R}$, charged under $Z_2^\tau$ and vector-like under the full gauge group. The Yukawa Lagrangian relevant to the $\tau$ sector is then:
\begin{equation}
- \mathcal{L}_Y^\tau = y_L\, \overline{\ell}_L^3 \E_R\, \S + M\, \overline{\E}_L \E_R + y_R\, \overline{\E}_L H\, e_R^3 + \text{h.c.} \, .
\end{equation}
Assuming $M$ is sufficiently large, the heavy FN mediator $\E$ can be integrated out, yielding the effective operator:
\begin{equation}
- \mathcal{L}_Y^\tau = - \frac{y_L y_R}{M}\, \overline{\ell}_L^3 H\, e_R^3\, \S + \text{h.c.} \, , 
\end{equation}
which generates the $\tau$ lepton mass (upon removing an overall sign via a chiral transformation):
\begin{equation}
m_\tau = \frac{y_L y_R}{2} \frac{v v_X}{M} = 1.78\, \text{GeV} \times y_L y_R \left( \frac{v_X/M}{0.014} \right) \,.
\end{equation}
In this limit, the effective IR $X$-current charges are simply $\alpha_L^\tau = 1$ and $\alpha_R^\tau = 0$. 
The prediction of this model is shown in red in the right panel of 
Fig.~\ref{fig:plot}.

Specifying the anomalon mass in \eq{eq:anomalonmass} 
to the case $n=2$ and $\alpha_L^\tau = 1$ we obtain 
\begin{align}
m_\L &= \frac{y_{\L}}{\sqrt{2}} 120 \, \text{GeV} \( \frac{m_X}{2.1 \, \text{GeV}} \) \( \frac{0.018}{{g_X}} \) 
\,, 
\end{align}
normalized to the best fit point $(m_X, g_X) = (2.1 \, \text{GeV}, 0.018)$.
The requirement of perturbative unitarity allows one to infer an upper bound on the Yukawa coupling,   $y_\L\leq\sqrt{4\pi}$~\cite{Allwicher:2021rtd}, thus $m_\L\leq290$ GeV.
Furthermore, at the best fit point one finds
\begin{align}
{\mathcal B}(Z\to\gamma X)_\text{best-fit}&=1.9\times10^{-7} \, .
\end{align}
The latter is sizeable, but still compatible with the LEP bound \cite{L3:1997exg}
in \eq{eq:LEPbound}. 
Hence, a future collider such as FCC-ee~\cite{FCC:2018evy} operating at the $Z$-pole could provide a test of this hypothesis.

\subsection{$L = -4 R$ model ($n=3$)}
\label{sec:Lm4Rmodel}

\begin{table}[!ht]
	\centering
	\begin{tabular}{|c|c|c|c|c|c|}
	\hline
	Field & Lorentz & $\SU(3)_C$ & $\SU(2)_L$ & $\U(1)_Y$ & $\U(1)_{X}$ \\ 
	\hline 
	$q^{i}_L$ & $(\tfrac{1}{2},0)$ & 3 & 2 & 1/6 & 0 \\ 
	$u^{i}_R$ & $(0,\tfrac{1}{2})$ & 3 & 1 & 2/3 & 0 \\ 
	$d^{i}_R$ & $(0,\tfrac{1}{2})$ & 3 & 1 & $-1/3$ & 0 \\ 
	$\ell^{1,2}_L$ & $(\tfrac{1}{2},0)$ & 1 & 2 & $-1/2$ & $0$ \\ 
	$e^{1,2}_R$ & $(0,\tfrac{1}{2})$ & 1 & 1 & $-1$ & $0$ \\ 
    $\ell^{3}_L$ & $(\tfrac{1}{2},0)$ & 1 & 2 & $-1/2$ & $4$ \\ 
	$e^{3}_R$ & $(0,\tfrac{1}{2})$ & 1 & 1 & $-1$ & $-1$ \\ 
	$H$ & $(0,0)$ & 1 & 2 & 1/2 & $0$ \\
	\hline
	$\L_L$ & $(\tfrac{1}{2},0)$ & 1 & 3 & 1 & $-3$ \\
	$\L_R$ & $(0,\tfrac{1}{2})$ & 1 & 3 & 1 & $-2$ \\
	$\N_R^{1}$ & $(0,\tfrac{1}{2})$ & 1 & 1 & $0$ & $2$ \\
    $\N_R^{2}$ & $(0,\tfrac{1}{2})$ & 1 & 1 & $0$ & $4$ \\
    \hline
    $\E_L$ & $(\tfrac{1}{2},0)$ & 1 & 1 & $-1$ & $4$ \\
	$\E_R$ & $(0,\tfrac{1}{2})$ & 1 & 1 & $-1$ & $4$ \\
    \hline
    $\S_1$ & $(0,0)$ & 1 & 1 & 0 & $1$ \\  
	$\S_5$ & $(0,0)$ & 1 & 1 & 0 & $5$ \\  
    \hline 
	\end{tabular}	
	\caption{\label{tab:Lm4Rmodel} 
	Field content of the $L = -4 R$ model.}
\end{table}

In this model gauge anomalies are canceled by an EW triplet ($n=3$) and two SM-singlet fields. We also introduce a fermion field, 
$\E$, vector-like under the full gauge group, and a new scalar field $\S_5$ to give mass to the $\tau$ lepton through a FN mechanism.
We show the field content of the model in \Table{tab:Lm4Rmodel}. 

By adding a proper term in the scalar potential, 
$\Delta V (H, \S_1, \S_5)$,\footnote{It is worth noting that, in the presence of only renormalizable operators, a massless Goldstone boson arises at the tree level. Its mass can be lifted either through loop corrections or by introducing a non-hermitian effective operator, such as $\S^*_5 \S^5_1$, in the scalar potential.} 
the following VEV configurations 
are generated
\beq 
\vev{H} = \frac{1}{\sqrt{2}} 
\begin{pmatrix}
0 \\ v
\end{pmatrix} \, , \ \ 
\vev{\S_1} = \frac{V_1}{\sqrt{2}} \, , \ \ 
\vev{\S_5} = \frac{V_5}{\sqrt{2}} \, ,
\eeq
with $v \approx 246$ GeV and $V_{1,5}$ being the order parameters of $\U(1)_X$ breaking. 
The latter are responsible for the mass of the $\U(1)_X$ gauge boson, that is
\begin{align} 
\label{eq:massX}
m_{X} &= g_X \sqrt{ V_1^2 + 25 V_5^2 } \equiv g_X v_X \, ,
\end{align}
where $g_X$ is the $\U(1)_X$ gauge coupling entering the covariant derivative, 
\emph{i.e.}~$D^\mu \S_{1,5} \equiv (\partial^\mu + i g_X \X_{\S_{1,5}} X^{\mu}) \S_{1,5}$. 

The renormalizable Yukawa Lagrangian 
involving uncolored fields reads ($a,b=1,2$)
\begin{align}
\label{eq:LYukALL}
- \mathcal{L}_{Y} &= 
y_\L \bar{\L}_{L} \L_{R} \S^{*}_1 
+ M \overline{E}_L E_R + y_{ab} \overline{\ell}_L^a e^b_R H \nonumber \\ 
&+ y_L \overline{\ell}_L^3 E_R H + y_R S_5 \overline{E}_L e_R^3 + \text{h.c.} \, .
\end{align}
The accidental global symmetry left unbroken 
by $\mathcal{L}_{Y}$
is 
$\U(1)_1 \times \U(1)_2 \times \U(1)_{3+E} \times \U(1)_{\L}$,
where $U(1)_{3+E}$ can be identified as the generalized $\tau$ number and $U(1)_{\L}$ is the anomalon number.

To identify the $\tau$ lepton, we introduce the mass matrix $\M_E$, defined via
\begin{equation}
\left( 
\begin{array}{cc}
\overline{e}_L^3 & \overline{\E}_L 
\end{array}
\right)
\left( 
\begin{array}{cc}
0 & \frac{1}{\sqrt{2}} y_L v \\
\frac{1}{\sqrt{2}} y_R V_5 & M
\end{array}
\right)
\left( 
\begin{array}{c}
e_R^3 \\
\E_R
\end{array}
\right) \, , 
\end{equation} 
where $e^3_L$ denotes the charged component of the doublet $\ell_L^3$.
In particular, from 
\beq 
\M_E \M_E^T = 
\left(
\begin{array}{cc}
 \frac{1}{2} y_L^2 v^2 & \frac{1}{\sqrt{2}} M y_L v \\
 \frac{1}{\sqrt{2}} M y_L v & M^2 + \frac{1}{2} y_R^2 V_5^2  \\
\end{array}
\right) \, ,
\eeq
and 
\beq 
\M_E^T \M_E = 
\left(
\begin{array}{cc}
 \frac{1}{2} y_R^2 V_5^2 & \frac{1}{\sqrt{2}} M y_R V_5 \\
 \frac{1}{\sqrt{2}} M y_R V_5  & M^2+\frac{1}{2} y_L^2v^2  \\
\end{array}
\right) \, , 
\eeq
we obtain the mixing angles (taking for simplicity real 
parameters)
\begin{align}
\tan 2\theta_L &=\frac{\sqrt{2} M y_L v}{\frac{1}{2} y_L^2 v^2 - M^2 - \frac{1}{2} y_R^2 V_5^2} \nonumber \\ 
&\approx - \sqrt{2} y_L v / M 
\, , \\ 
\tan 2\theta_R &=\frac{\sqrt{2} M y_R V_5}{\frac{1}{2} y_R^2 V_5^2 - M^2 - \frac{1}{2} y_L^2v^2} \nonumber \\ 
&\approx - \sqrt{2} y_R V_5 / M 
\, ,
\end{align}
where the approximation holds in the 
$M \gg V_5, v$ limit. 
We hence obtain the $\tau$ chiral components in terms of gauge eigenstates: 
\begin{align}
\tau_{L,R} &= \cos \theta_{L,R} e^3_{L,R} +  \sin \theta_{L,R} E_{L,R} \, , \\
\T_{L,R} &= -\sin \theta_{L,R} e^3_{L,R} +  \sin \theta_{L,R} E_{L,R} \, .
\end{align}
Projecting \eq{eq:LYukALL} on the mass eigenstates, 
we obtain 
\begin{align}
\label{eq:mtau}
m_\tau &\approx \frac{y_L y_R}{2} v \frac{V_5}{M} \nonumber \\
&= 1.78 \, \text{GeV} \, y_L y_R \( \frac{V_5/M}{0.014} \) \, , \\ 
m_\T &\approx M \, , 
\end{align}
implying that $\theta_R$ must be small to account for the $\tau$-lepton mass.

Note that this setup induces a deviation from unitarity in the PMNS matrix, due to the modified structure of the third-generation lepton doublet:
\begin{equation}
\ell_L^3 =
\left(
\begin{array}{c}
\cos \theta_L  \tau_L \\
\nu_{\tau}
\end{array}
\right) \, ,
\end{equation}
which implies that $\theta_L$ is constrained to be close to zero at the percent level, consistent with neutrino oscillation data (see \emph{e.g.}~\cite{Forero:2021azc}). 

Hence, in the limit of negligible $\theta_{L,R}$, the effective couplings of the IR $X$-current are given by
\begin{align}
\alpha_L^\tau&=4 \, , \hspace{3mm} \alpha_R^\tau=-1 + 5 \sin^2 \theta_R \approx-1 \, .
\end{align}
In the right panel of Fig.~\ref{fig:plot}, we display the prediction of this model in blue.
Given the best fit point $(m_X, g_X) = (2.1 \, \text{GeV}, 0.0042)$, this yields $v_X \approx 467$ GeV,
along with the predicted rate
for the $Z \to \gamma X$ decay
\begin{align}
{\mathcal B}(Z\to\gamma X)_\text{best-fit}&=1.2\times10^{-7} \, .
\end{align}
The anomalons pick up the mass term  
\begin{align} 
m_{\L} &=  
\frac{y_{\L}}{\sqrt{2}} V_1 
\approx \frac{y_{\L}}{\sqrt{2}} \frac{m_X}{g_X} \nonumber \\
&= \frac{y_{\L}}{\sqrt{2}} 
\, 500 \, \text{GeV} \, 
\( \frac{m_X}{2.1 \, \text{GeV}} \)
\( \frac{0.0046}{g_X} \)
\, , 
\end{align}
where we assumed $v_X \approx V_1 \gg V_5$. 
The perturbative unitarity bound on the Yukawa coupling reads in this case $y_\L\leq\sqrt{8\pi/3}$~\cite{Allwicher:2021rtd}, thus $m_\L\leq1.0$ TeV.

The SM-singlet states $\N_R^{1,2}$ 
pick-up a mass term from the $d=6$ operators
\begin{equation}
\mathcal{L}_{\N} = \frac{\lambda_1}{\Lambda^2} \S \S^*_5 \N^1_{R} \N^1_{R} 
+\frac{\lambda_2}{\Lambda^2} \S^* \S_5^* \N^1_{R} \N^2_{R} + \text{h.c.} \, . 
\end{equation}
While their masses are controlled by the cutoff scale $\Lambda$, these SM-singlet states do not mix with SM fermions and therefore play only a marginal role in the low-energy phenomenology.

A final comment about neutrino masses is in order. 
We work in the limit of massless neutrinos. 
However, some extra dynamics is required in order to generate neutrino masses and a non-trivial PMNS. 
This dynamics can be pushed up to scales $\gg$ TeV, and therefore it decouples from the relevant phenomenology we 
focus on.

\section{Conclusions}
\label{sec:concl}

We have explored a class of models where a light spin-1 boson couples to a chiral component of the $\tau$-lepton current via a $\U(1)_X$ gauge symmetry. The resulting low-energy EFT features anomaly-induced WZ interactions with EW gauge bosons, leading to distinct signatures in flavor and EW observables.

As a concrete application, we have shown that such a boson with mass $m_X \simeq 2.1\,\text{GeV}$ can account for the recent $3\sigma$ excess observed in $B^+ \to K^+ E_{\rm miss}$ at Belle II, while remaining compatible with current bounds from $Z \to \gamma E_{\rm miss}$ and direct searches for anomalons. The dominant decay mode of $X$ is into neutrinos, consistent with the missing energy signature, and its coupling to quarks arises from SM flavor violation dressing the $X W \partial W$ vertex.
In fact, our work offers a first concrete setup that accounts for the $B^+ \to K^+ E_{\rm miss}$ excess within the framework of MFV.

We constructed a minimal anomaly-free renormalizable UV completion based on gauging chiral $\tau$-lepton number and introduced two explicit realizations.  These scenarios offer rich and predictive phenomenology, with additional experimental signatures that could be tested in future high-energy and intensity frontier experiments.

In the first model ($L$-only), we gauged the left component of $\tau$-lepton number. The model features a minimal set of couplings and fields and predicts the presence of EW anomalons very close to the current experimental bounds. However, from a theoretical standpoint, this model requires ad-hoc symmetries in the anomalon and lepton sectors to avoid lepton-flavor violation.

While the phenomenological predictions of the two models are similar, the second model, referred to as $L = -4R$, is theoretically more appealing. The $\tau$ sector is automatically selected by the gauge symmetry, and the accidental symmetries of the SM are automatically preserved.

Last, but not least, our work also aims to emphasize the importance of providing a UV completion for scenarios in which light degrees of freedom beyond the SM appear at low energies.

\section*{Acknowledgments}
We thank Jernej Kamenik for useful discussions 
at the early stages of this work. We also thank Gabriele Levati 
for helpful comments on an earlier version of the paper.
LDL and MN are supported
by the 
European Union -- Next Generation EU and
by the Italian Ministry of University and Research (MUR) 
via the PRIN 2022 project n.~2022K4B58X -- AxionOrigins.
The work of CT has received funding from the French ANR, under contracts ANR-19-CE31-0016 (`GammaRare') and ANR-23-CE31-0018 (`InvISYble'), that he gratefully acknowledges.

\bibliographystyle{apsrev4-1.bst}
\bibliography{bibliography}

\end{document}